\documentclass[aps,prb,reprint,groupedaddress,showpacs]{revtex4-1}
\usepackage{graphicx}
\usepackage{color}
\usepackage{amssymb}
\usepackage{amsmath}
\usepackage{bm}

\begin{document}

\title{Phonon decay in silicon nanocrystals}
\author{A.\,A.~Prokofiev}\email{lxpro@mail.ioffe.ru}
\author{A.\,N.~Poddubny}
\author{I.\,N.~Yassievich}
\affiliation{Ioffe Physical-Technical Institute of RAS, 
194021 St.~Petersburg, Russia}

\date{\today}

\begin{abstract}
	The decay of the optical phonon into the two phonons of smaller energy is calculated for Si nanocrystals. The rate of the process is in the  range of 1 to 10~ps.  Such anharmonic phonon decay  may 
control the energy relaxation rate of  excited carriers in 
Si/SiO$_2$ nanocrystals. Relevance of the phonon decay to the experimentally observed  hot carrier photoluminescence is discussed.
\end{abstract}


\maketitle

\section{Introduction}
\label{sec:intro}
Silicon nanocrystals form a rapidly developing area of research with a number of promising applications for optoelectronics  and photovoltaics. 
Rapt attention is now attracted to such processes as space-separated quantum cutting\cite{Timmerman2008,Trinh:2012:NatPhot,Govoni:2012:NaturePhot} and  hot photoluminescence,\cite{deBoerNatureNanotech2010} which intrinsically involve hot confined carriers, optically pumped inside the nanocrystal.
This raises the demand to clarify the mechanisms, controlling the  hot carrier lifetime.

\begin{figure}[t]
  \centering
  \includegraphics[width=8cm]{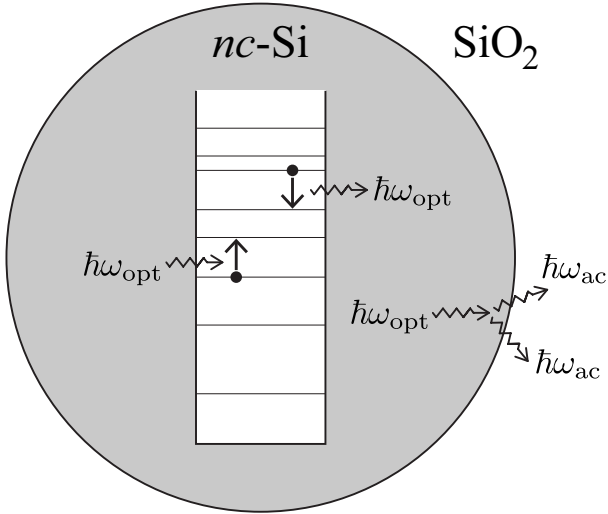}
  \caption{\label{fig:scheme} The schematical representation of the suggested relaxation mechanism. 
Confined carriers in Si nanocrystal emit optical phonons (as well as absorb them) very fast.
Optical phonons decay into acoustic ones, which can leave the nanocrystal taking the energy away.}
\end{figure}

The general understanding is that the energy relaxation of hot carriers  is determined by the emission of optical phonons.\cite{Delerue2004,Govoni:2012:NaturePhot} For lowest excited states 
this process can be slowed down  in nanocrystal with respect to that in the bulk due to the energy spectrum discretization.\cite{MoskalenkoPRB2012} However, this argument does not hold for hot excited states, which require completely different consideration. We will show below, using the atomistic calculations of carrier and phonon spectra, that the rate of energy relaxation due to optical phonon emission may reach the values up to $0.1$~eV/fs. Such ultrafast relaxation is in clear contradiction to the various experimental manifestations of the hot carriers presence in the nanocrystal during $1\div 100~$ps after the pump pulse.\cite{deBoerNatureNanotech2010,WietekePRB2012rapid} To resolve the paradox we propose the following scenario, see Fig.~\ref{fig:scheme} (i) Electrons and holes emit optical phonons very fast. (ii) These optical phonons do not escape the nanocrystal due to their low group velocity and  differences between the phonon spectra in Si and SiO$_2$.\cite{wei1993} Instead, they are quickly reabsorbed by the confined carriers, so that the total energy of the electron-phonon system is conserved.  (iii) Phonon recycling process stops after the only after the anharmonic decay of high-energy phonons into the lower energy ones. This low energy acoustic-like phonons can relatively easy leave the nanocrystal, so the energy of the system ``carriers+phonons'' is dissipated. Thus, the energy relaxation rate is governed not by the phonon emission rate, but by the optical phonon decay rate.
Similar mechanisms are known in literature and important, for instance, for GaN devices.\cite{Khurgin:2007:APL}

The paper is organized as follows. Calculation approach for the phonon modes is outlined in Sec.~\ref{sec:modes}. Phonon emission and phonon decay processes are considered in Sec.~\ref{sec:relax} and Sec.~\ref{sec:decay}. Main paper results are summarized in Sec.~\ref{sec:concl}.

\section{Phonon modes of the Si nanocrystal}\label{sec:modes}
\label{sec:calc}
\begin{figure}[t]
  \centering
  \includegraphics[width=8cm]{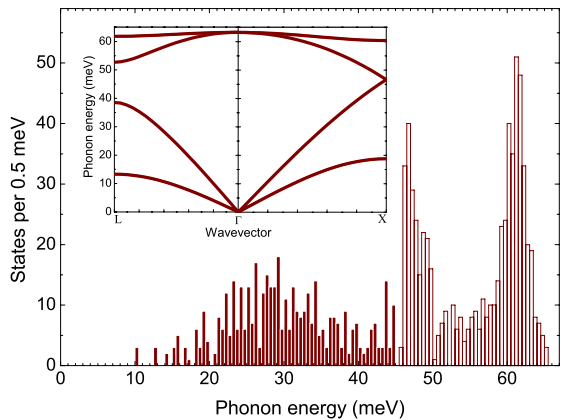}
  \caption{\label{fig:dos} Calculated density of phonon states in Si nanocrystal with diameter 2.5~nm.
Open bars correspond to ``optical'' phonons (with energies above the phonon energy gap).
The inset shows phonon dispersion of bulk silicon calculated in the same model.
}
\end{figure}

Dispersion of phonons in bulk silicon can been described within various Keating-type models,\cite{Keating:1966:PR,vanderbilt1989,Rueker:1995:PRB}
where the atoms are effectively treated as balls connected by springs.
 In our approach we adopt the simplest two-parameter Keating model from Ref.~\onlinecite{Rueker:1995:PRB}, which allows straightforward generalization for anharmonicity. According to this model, the strain energy is given by
\begin{equation}
	W 
= \sum_{i,j} {\alpha \over a_0^2}
	\left[
		\Delta  \left( r_{ij}^2 \right)
	\right]^2
+ \sum_{i,j,k\neq j} {\beta \over a_0^2}
	\left[
		\Delta 
			\left(
				\mathbf{r}_{ij} \cdot \mathbf{r}_{ik}
			\right)
	\right]^2\:,
\label{eq:strain_harmonic}
\end{equation}
where $a_0$ is the lattice constant, 
and $\Delta$ denotes the change  to the unstrained lattice:
$\Delta  ( r^2 ) = r^2 - r_0^2$, $\Delta (\mathbf{r}_{ij} \cdot \mathbf{r}_{ik}) = (\mathbf{r}_{ij} \cdot \mathbf{r}_{ik})-(\mathbf{r}_{ij}^0 \cdot \mathbf{r}_{ik}^0)$.
The two constants $\alpha$ and $\beta$ describe the bond-stretching and bond-bending forces. 
In the harmonic approximation the Hamiltonian  of the vibrational system assumes the form
\begin{equation}
	\hat{H} 
		= \sum_i \frac{\hat{p}^2}{2M_{\mathrm{Si}}} 
		+ \sum_{i,j} V_{ij} x_i x_j,
\label{eq:harmonic_H}
\end{equation}
where $M_{\mathrm{Si}}$ is the mass of silicon atom, 
$\hat{p}$ is the momentum operator,
and
\begin{equation}
	V_{ij} = \frac{\partial^2 W}{\partial x_i \partial x_j}\:.
\label{V_ij}
\end{equation}
The phonon modes in the nanocrystal have been calculated using the clamped boundary conditions:  the atoms displacements of all the atoms outside the  nanocrystal were set to zero. Relaxation of the nanocrystal lattice has been neglected for simplicity. 
The values of constants $\alpha$ and $\beta$ from  Ref.~\onlinecite{Rueker:1995:PRB} have been used.
The calculated density of phonon states in Si nanocrystal with diameter 2.5~nm is shown in Fig.~\ref{fig:dos}.
The figure demonstrates that the phonon modes can be divided into two distinct groups. 
First group corresponds to the acoustic-like modes with the energy 
$\hbar \omega\lesssim 40$~meV (filled bars). 
Size  quantization of the phonons explains the absence of the modes with the  energies less 
than 12~meV. 
Second group of the  modes with $\hbar\omega\gtrsim 40~$meV corresponds to optical-like phonons and is manifested as two distinct peaks in the density of phonon states (outlined bars). These results are in qualitative agreement with Ref.~\onlinecite{Valentin2008} where a detailed theoretical study of phonon modes in Si nanocrystals is presented.

\section{Phonon emission processes}
\label{sec:relax}
Phonon emission processes, usually responsible for energy relaxation, can be considered as follows.  First, the energy spectra of confined carriers and phonons in a nanocrystal are calculated separately. Second, the interaction of carriers with phonons is analyzed using the fact that the distortion of the atomic positions due to the vibrations changes the matrix elements of the tight-binding Hamiltonian for the confined carriers.

We have used nearest-neighbor $sp^3d^5s^*$ tight-binding method to calculate the probability of single-phonon transitions of excited electrons and holes in Si nanocrystals.\cite{jancu1998} The dangling bonds at the nanocrystal surface were ``passivated'' with hydrogen atoms, the parameters were taken from Ref.~\onlinecite{hill1996}. The relaxation of the nanocrystal lattice from the bulk has been neglected. Calculated densities of states of confined carriers and more details of the model can be found in Ref.~\onlinecite{poddubny2010apl}.

The emission rate of the single optical phonon has been calculated according to the Fermi Golden rule as 
\begin{equation}\label{eq:tau}
 \frac{1}{\tau_i}=\frac{2\pi}{\hbar}\sum\limits_{\nu}\sum\limits_f\langle f|H^{(\nu)}_{\mathrm{e-ph}}|i\rangle^2\delta(E_i-E_f-\hbar \omega_{\nu})\:,
\end{equation}
where $E_i$ and $E_f$ are the energies of initial and final state and the summation runs over the phonon states with the energies $\hbar\omega_{\nu}$.
In numerical calculation the $\delta$-function, accounting for energy conservation, has replaced by a Lorentzian with HWHM equal to $1$~meV. The Hamiltonian of carrier interaction $H_{\mathrm{e-ph}}$ with the given phonon mode $\nu$ has been found by taking the derivative of the tight-binding Hamiltonian over displacement:
\begin{equation}
	H_{\mathrm{e-ph}}^{(\nu)}
	=
	\sum_i \frac{\partial H_{\mathrm{e}}}{\partial{\bm r}_{ij}} \delta \bm r_{ij}^{(\nu)}\:.
\label{eq:H_e-ph}
\end{equation}
where 
$\delta \bm r^{(\nu)}_{ij}=\bm u^{(\nu)}_i-\bm u^{(\nu)}_j$ are the atomic displacements, corresponding to the  phonon mode $\nu$. The dependence of the matrix elements of the tight-binding Hamiltonian on the interatomic distance has been taken from Ref.~\cite{jancu1998}. 
 Energy relaxation rate for a given state $i$ is obtained by multiplying each term in Eq.~\eqref{eq:tau} by 
$\hbar\omega_\nu$.

\begin{figure}[t]
  \centering
  \includegraphics[width=8cm]{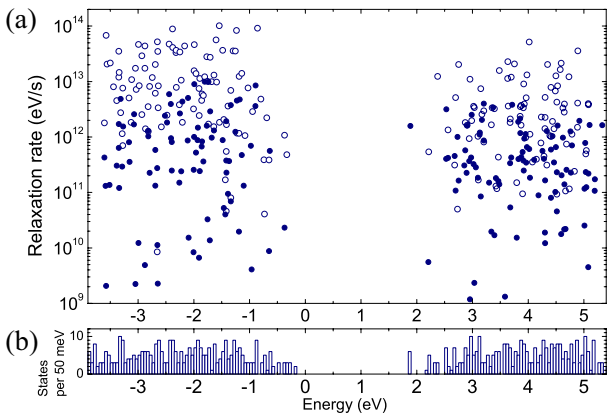}
  \caption{\label{fig:relax} Relaxation rate of electrons and holes (negative energies) in Si$_{363}$ nanocrystal ($d \approx 2.5$~nm). 
Each point corresponds to single-phonon transition rate from the state with given energy,
related to the energy of phonon.
Open symbols indicate the transitions involving phonons with energy higher than 40~meV.
The histogram in panel (b) referres to the density of electronic states in the nanocrystal.
}
\end{figure}

Fig.~\ref{fig:relax} shows calculated energy relaxation rates of electrons and holes in 2.5~nm silicon nanocrystal due to single-phonon transitions from all possible excited states.
The figure demonstrates that the energy relaxation rate can be quite fast and may reach the values $\sim 10^{14}$~eV/s, i.e. the energy of one eV could be lost within the tens of femtoseconds. Majority of the transitions correspond to a somewhat slower relaxation rate within the range  $\sim 10^{12}\div 10^{13}$~eV/s. 
These new relaxation rates are significantly higher than those reported in our previous work,\cite{poddubny2010apl} the discrepancy is due to the technical mistake in Ref.~\onlinecite{poddubny2010apl}.

Noteworthy, the fastest relaxation processes are due to the emission of optical phonons, see open symbols in  Fig.~\ref{fig:relax}.
Thus, hot carriers in nanocrystal may generate a large population of  optical phonons. If these phonons do not leave the nanocrystal fast enough, they can be reabsorbed by the carriers. This means that the phonon emission process itself does not lead to the dissipation of the total energy in the system. The energy relaxation is controlled by another processes, such as anharmonic phonon decay of optical phonons into acoustic ones, analyzed below.\cite{Ridley:1996:JPCM,Khurgin:2007:APL,Gurevich:1980:book}

 \section{Phonon decay processes}
 \label{sec:decay}
Phonon anharmonicity may be accounted for by adding the terms
in the strain energy Eq.~\eqref{eq:strain_harmonic}, cubic in the atomic displacement. This is done by including the dependence of the force constants  $\alpha$ and  $\beta$ in Eq.~(\ref{eq:strain_harmonic}) on the bond lengths $r_{jk}$:\cite{Rueker:1995:PRB}
\begin{eqnarray}
	\alpha
	&=& \alpha_0 
		\left({r^0_{jk} \over r_{jk}}\right)^{4},
\\
	\beta
	&=& \beta_0 
		\left({r^0_{jk} \over r_{jk}}\right)^{7/2}
		\left({r^0_{kl\vphantom{j}} \over r_{kl\vphantom{j}}}\right)^{7/2}.
\label{eq:alpha-beta}
\end{eqnarray}
Here, the values $r_{jk}^0$ correspond to the undistorted lattice,
and $\alpha_0, \beta_0$ are the values of constants $\alpha,\beta$ used in harmonic approximation.
The anharmonic part of the strain energy, cubic in displacement, can be then  reduced to
\begin{equation}
	\hat{H}_{\mathrm{anh}} = \frac{1}{6} \sum_{ijk} V_{ijk} x_i x_j x_k.
\label{eq:H3}
\end{equation}
Using the second-quantization formalism, we rewrite the harmonic part of the phonon Hamiltonian as 
\begin{equation}
	\hat{H} = \sum_\nu \hbar\omega_\nu \left(b_\nu^\dag b_\nu + {1\over 2}\right)\:.
\label{eq:harmonic_H_bb}
 \end{equation}
The terms of the anharmonic Hamiltonian responsible for the decay of the  phonon mode $\nu$ into the modes $\mu$ and $\kappa$ reads
\begin{equation}
	\hat{H}_{\mathrm{anh}} 
		= \sum_{\nu\mu\kappa} 
				A_{\nu\mu\kappa} 
 				b_\nu^{\vphantom{\dag}} b_\mu^\dag b_\kappa^\dag\:,
\label{eq:H3_bbb}
\end{equation}
where
\begin{equation}
	A_{\nu\mu\kappa} 
	= {1\over 6} 
		\sum_{ijk} 
		\sqrt{\frac{\hbar^3}{(2 M_{\rm Si})^3 \omega_\nu \omega_\mu \omega_\kappa}} 
		V_{ijk} 
		u_i^\nu u_j^\mu u_k^\kappa\:.
\label{A_abg_def}
\end{equation}
The decay rate of the phonon mode $\alpha$  reads
\begin{equation}
	{1 \over \tau_\nu}
	=
	{2 \pi \over \hbar}
	\sum_{\mu,\kappa}
		\left|
			3 A_{\nu\mu\kappa}
		\right|^2
		\delta
			\left(
					\hbar\omega_\nu - \hbar\omega_\mu - \hbar\omega_\kappa
			\right).
\label{eq:tau_final}
\end{equation}
In numerical calculation the $\delta$-function, accounting for energy conservation, has been replaced by a Lorentzian with HWHM equal to $0.15$~meV.

\begin{figure}[t]
  \centering
  \includegraphics[width=8cm]{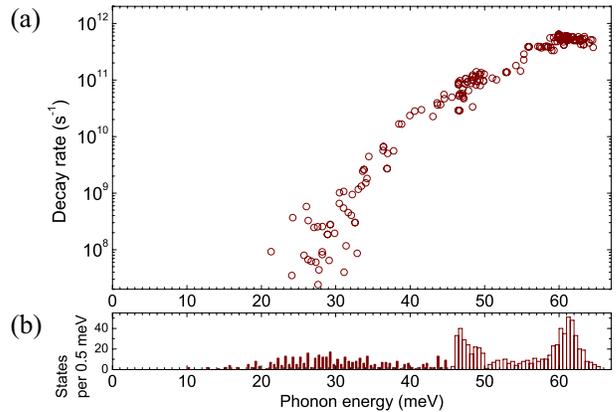}
  \caption{\label{fig:decay} Phonon decay rate in Si$_{363}$ nanocrystal ($d \approx 2.5$~nm).
The density of phonon states in the nanocrystal is shown in panel (b) for reference.
}
\end{figure}

The calculated rates of phonon decay for all the modes of Si$_{363}$ nanocrystal with the diameter close to 2.5~nm are shown in Fig.~\ref{fig:decay}. The figure demostrates that the decay time monotonously decreases with the phonon energy and reach the value $\sim 1~$ps. This 
growth of the phonon decay rate with phonon energy reflects the increasing density of final states. Our analysis indicates, that the phonon emission rate weakly depends on the nanocrystal  size. It may be estimated using the following simple argument.
The element of cubic phonon-phonon interaction is approximately given by
the product of the phonon energy and the ratio of the 
vibration amplitude to the lattice constant,
\begin{equation}
	A_{\nu\mu\kappa} \propto \hbar\omega_\nu {u_\nu \over a_0}\:.
\label{eq:Aabg_simple}
\end{equation}
The value of $u_\nu$ can be found from the normalization condition $M_{\rm Si}u_\nu^2=\hbar\omega_\nu$. Finally, the density of final phonon states
in Eq.~\eqref{eq:tau_final} can be estimated as $1/\omega_{\nu}$. This yields $(\tau_\nu)^{-1} \approx 10^{11}$~s$^{-1}$, in agreement with Fig.~\ref{fig:decay}.

\section{Conclusion}
\label{sec:concl}
Theory of  phonon emission and phonon decay processes in silicon nanocrystals has been developed. The calculation has been carried out using the empirical
$sp^3d^5s^*$ model for the states of confined carriers and electrons combined with the anharmonic Keating model for the vibrations of the nanocrystal atoms.

The calculations yield very short times of optical phonon emission, reaching tens of femtoseconds for high energy optical-like phonons. This is an indication for a strong electron-phonon interaction. Hence, emitted  phonons can be quickly reabsorbed, and the total energy of the system ``electrons+phonons'' is conserved. The energy dissipation rate is then controlled by the anharmonic decay of the high-energy phonons into several low-energy acoustic-like ones, which then leave the nanocrystal.

Calculated phonon lifetimes due to three-phonon decay processes are on the order of tens of picoseconds and weakly depend on the  nanocrystal size. Such  decay can govern the total energy relaxation rate of hot carriers in silicon nanocrystal. The value of the rate corresponds to the lifetime of hot carriers 
determined from the decay of the hot photoluminescence  in Si/SiO$_2$ nanocrystals.\cite{deBoerNatureNanotech2010}
\begin{acknowledgments}
The work has been supported by RFBR.
Useful discussions with T.~Gregorkiewicz and K.~Dohnalov\'{a} are gratefully acknowledged.
\end{acknowledgments}

\bibliography{phonon_paper}

\end{document}